\documentclass[preprint,showpacs,preprintnumbers,amsmath,amssymb,12pt]{revtex4}
\usepackage{graphicx}
\usepackage{dcolumn}
\usepackage{bm}
\usepackage{amssymb}
\usepackage{amsmath}
\usepackage{latexsym}

\begin{document}

\title{EMISSION OF CHARGED PARTICLES FROM EXCITED COMPOUND NUCLEUS}
\author{Sh.A.Kalandarov$^{1,2}$}
\affiliation{$^{1}$Joint Institute for Nuclear Research, 141980 Dubna, Russia\\
$^{2}$Institute of Nuclear Physics,702132 Tashkent, Uzbekistan}

\begin{abstract}
The formation of excited compound nucleus (CN) and its statistical decay is investigated within the dinuclear system (DNS) model.The initial DNS is formed in the entrance channel when the projectile is captured by a target, and then the evolution of DNS in mass asymmetry coordinate leads to formation of the hot CN. The emission barriers for complex fragments were calculated within the DNS model by using the double folding procedure for the interaction potential. It is shown that cross sections for complex fragment emission become larger when excited CN is more neutron deficient. This approach gives also an opportunity to calculate the new neutron deficient isotopes production cross sections and can be applied to describe the hot fission of heavy systems.The model was tested by comparison of calculated results with  experimental datas for the  $^{3}$He+ $^{108}$Ag, $^{78,}$  $^{86}$Kr+ $^{12}$C,$^{63}$Cu+ $^{12}$C reactions.

\end{abstract}

\maketitle

\section{Introduction}

Complex fragment  emission($Z>2$) in intermediate energy nuclear
reactions has been a subject of both experimental and theoretical
interest for many years\cite{Lynch,Mor1}. The early studies of
this process identified two components: a fast, non-equilibrium component producing light
fragments at forward angles, and a relaxed component producing
fragments at all angles. Systematic studies of the relaxed
component at low bombarding energies demonstrated the compound
nucleus(CN) nature of the emission process \cite{Sobotka}.
We consider the formation of CN  and its statistical decay
within the dinuclear system(DNS) model\cite{DNS}.
The evolution of DNS in mass(charge) asymmetry coordinate
is responsible for the formation of CN and its decay.

\section{ CN formation and its decay}

Let's consider two colliding nuclei, with mass and charge numbers
$A_{P},Z_{P}$(projectile) and  $A_{T},Z_{T}$(target) with an
initial kinetic energy $E_{lab}$ of projectile. The first step in
formation of CN is the capture of the projectile by the target which
leads to formation of dinuclear system(DNS). The probability of capture process depends on
dissipation of kinetic energy of projectile, on angular momentum $J$ of the system and on mass(charge) asymmetry of
colliding nuclei. For a given projectile-target combination, there is some critical value
of $J$, in which the potential pocket disappears. At higher angular momentums than $J_{cr}$, projectile can not be captured. For the reactions which we consider here, all partial waves from $J=0$ to $J_{max}$ contribute to capture cross section. $J_{max}$ is determined from $min[J_{cr},J^{kin}_{max}]$, where $J^{kin}_{max}=(2\mu(E_{c.m.}-V_b))^{1/2}R_b$ with $V_b=V_N(R_b)+V_C(R_b)$, $V_N(R_b)$-nuclear potential,
$V_C(R_b)$ is a Coulomb potential, $R_b$ is  Coulomb barrier distance for the entrance channel.
The partial capture cross section is $\sigma_{cap}=\pi\lambdabar^{2}(2J+1)T(E_{c.m.},J),$
where $\lambdabar^{2}=\hbar^{2}/(2\mu E_{c.m.})$, $\mu$ is reduced
mass and $T(E_{c.m.},J)$ is a probability of crossing the Coulomb
barrier. For the energies well above the Coulomb barrier $T(E_{c.m.},J)=1$.

From the initial DNS configuration, system can evolve in three different ways: in a direction to complete fusion, to
quasifission or in a direction to the symmetric configuration.
The dynamics of the system in mass(charge) asymmetry coordinate
is determined by  driving potential\cite{driving} and by quasifission barrier $B_{qf}$ at each
configuration. For the asymmetric reactions, at zero angular momentum, the  quasifission barrier $B_{qf}$ and barrier for going to symmetric configuration $B^{sym}_{\eta}$ is very high relatively to fusion barrier $B_{\eta}$,  so the system will evolve with high probability to fusion. The complex fragment emission cross section given as
\begin{eqnarray}
\sigma_{e}(E_{ex},J)=\sum_{J=0}^{J=J_{max}}\sigma_{cap}(E_{c.m},J)P_{CN}(E_{c.m.},J)P_{e}(E_{ex},J),
\end{eqnarray}
where
\begin{eqnarray}
 P_{CN}=\frac{\exp(-B_{\eta}/T)}{\exp(-B_{\eta}/T)+\exp(-B_{qf}/T)+\exp(-B^{sym}_{\eta}/T)}
\end{eqnarray}
 is a probability of complete fusion which depends on competition between fusion and quasifission. Here we use Fermi-gas expression for the nuclear temperature $T$ with excitation energy of compound nucleus $E_{ex}=Q+E_{c.m.}-V_{rot}$. We use rigid body moment of inertia for calculation of rotational energy $V_{rot}$.

$$P_{e}(E_{ex},J)=\frac{W(i)\Gamma_i}{\sum_{i'}W(i') \Gamma_{i'}}
$$  in equation (1) is the emission probability of a given
particle from excited CN, $\Gamma_i$ is a decay width of a given channel, $W(i)$ is the normalized weight factor for given DNS configuration. Decay width is given by transition state formalism
as in Ref. \cite{Mor1}. We use Fermi-gas model expression for level densities.
The emission of complex fragments from excited CN follow two
steps: first this complex fragment must be formed and then overcome the emission barrier. The formation of complex fragment we understand as transition from CN configuration to
DNS configuration. Each DNS configuration has a weight factor which is of the form $\sim \exp(-U_i/T)$, where $U_i$ is the potential energy of DNS configuration at minima in relative distance between nuclei. The CN configuration has the weight factor $\sim \exp(-U_{CN}/T)$.  For the emission of light particles until alpha particles, we use binding energy+coulomb barrier as the emission barrier.
The excitation energy of given DNS is $E^{DNS}_{ex}=E_{ex}-U_{min}$,
where $U_{min}$ is the potential energy of DNS(determined with respect to CN energy) of given configuration at minima.
If $E_{ex}<U_{min}$, then the complex fragment can't be formed.
The formation of DNS, which lies in the right side of the
B.G. point in driving potential is possible only if excitation energy is larger than $U^{B.G.}_{min}$.
For the generation of cascade decay process of excited CN, we use Monte-Carlo technique. After each decay process, we recalculate the excitation energies of decay products as

$$
E^{1}_{ex}=(E^{DNS}_{ex}-(U_b-U_{min}))\frac{A_1}{A},
E^{2}_{ex}=(E^{DNS}_{ex}-(U_b-U_{min}))\frac{A_2}{A},
$$
where $A$ is the mass number of decaying CN, $U_b$ is potential energy at the barrier for a given DNS configuration.

\section{ Results of calculations}
\begin{figure}
\begin{minipage}[t]{7cm}
\begin{center}
\includegraphics[width=7cm,clip]{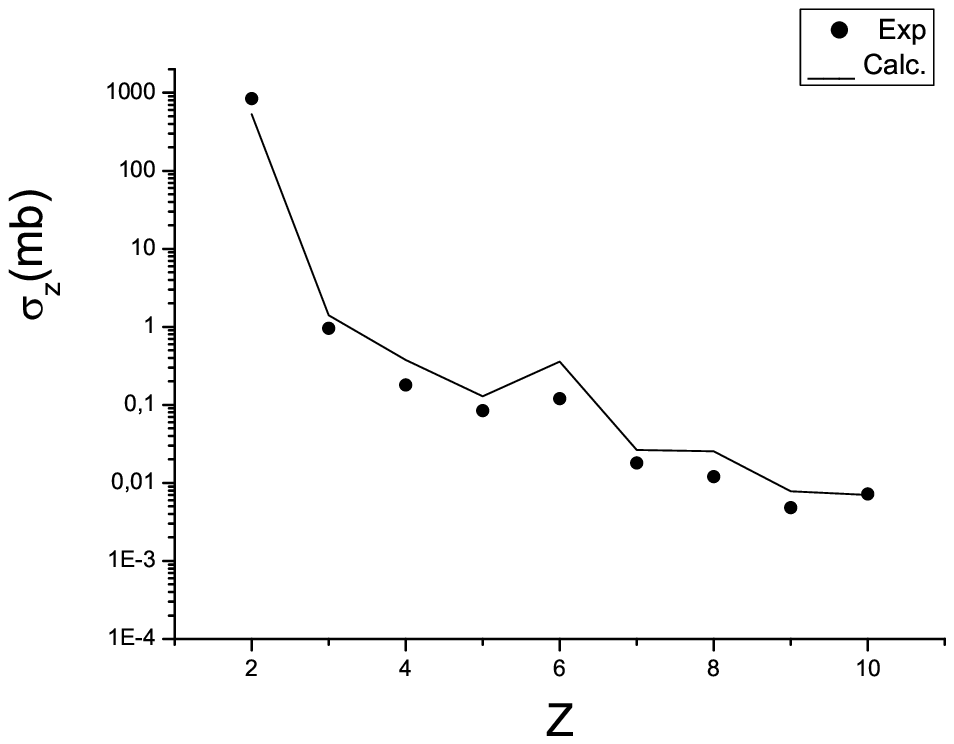}
\caption[Short caption for figure 1]{\label{HeAg} Comparison of calculated emission cross sections(line) and experimental data(point) for the reaction $^{3}$He+ $^{108}$Ag with initial bombarding energy  $E=30$MeV/nucleon.}
\end{center}
\end{minipage}
\hfill
\begin{minipage}[t]{7cm}
\begin{center}
\includegraphics[width=7cm,clip]{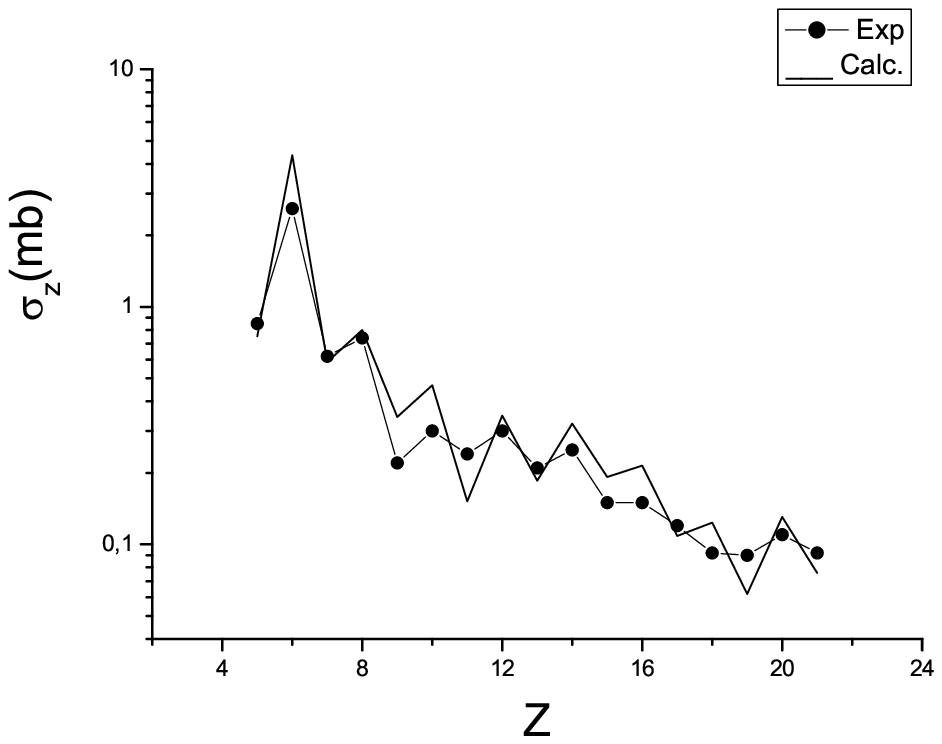}
\caption{\label{78Kr12C} Comparison of calculated emission cross sections for the reaction
$^{78}$Kr+ $^{12}$C with initial bombarding energy
$E=8.52$MeV/nucleon.}
\end{center}
\end{minipage}
\end{figure}

In the decay process, when CN decays into two nearly equal pieces, it produces
 always neutron rich nuclei, and in this cases if the neutron is
 unbound, then we just evaporate it from the nuclei without taking any excitation energy.
 First we checked the calculation results for the reaction
 $^3$He + $^{108}$Ag with $E_{lab}=90$ MeV leading to CN $^{111}$In, for which the experiment was performed by
 L.G.Sobotka et. al.\cite{Sobotk}. For asymmetric reactions
 the fusion barrier which is determined by driving potential is sufficiently lower than the quasifission barrier and
 barrier for going to symmetric configuration. In  this reaction  $J^{kin}_{max}=17 \hbar$ and $J_{cr}=15 \hbar$,
 and one should take $J_{max}=15 \hbar$. The calculated emission cross sections  are in good agreement with experimental data which is shown in  Fig. \ref{HeAg}.

\begin{figure}
\begin{minipage}[t]{7cm}
\begin{center}
\includegraphics[width=7cm,clip]{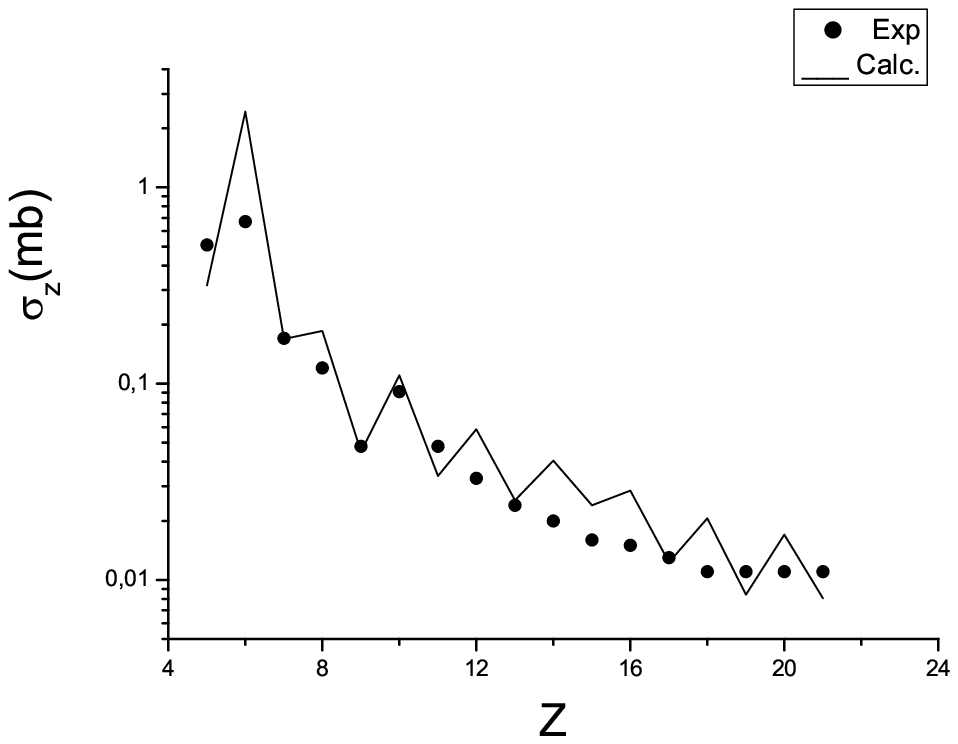}
\caption[Short caption for figure 1]{\label{86Kr12C} Comparison of calculated emission cross sections for the reaction
$^{86}$Kr+ $^{12}$C with initial bombarding energy $E=9.31$MeV/nucleon.}
\end{center}
\end{minipage}
\hfill
\begin{minipage}[t]{7cm}
\begin{center}
\includegraphics[width=7cm,clip]{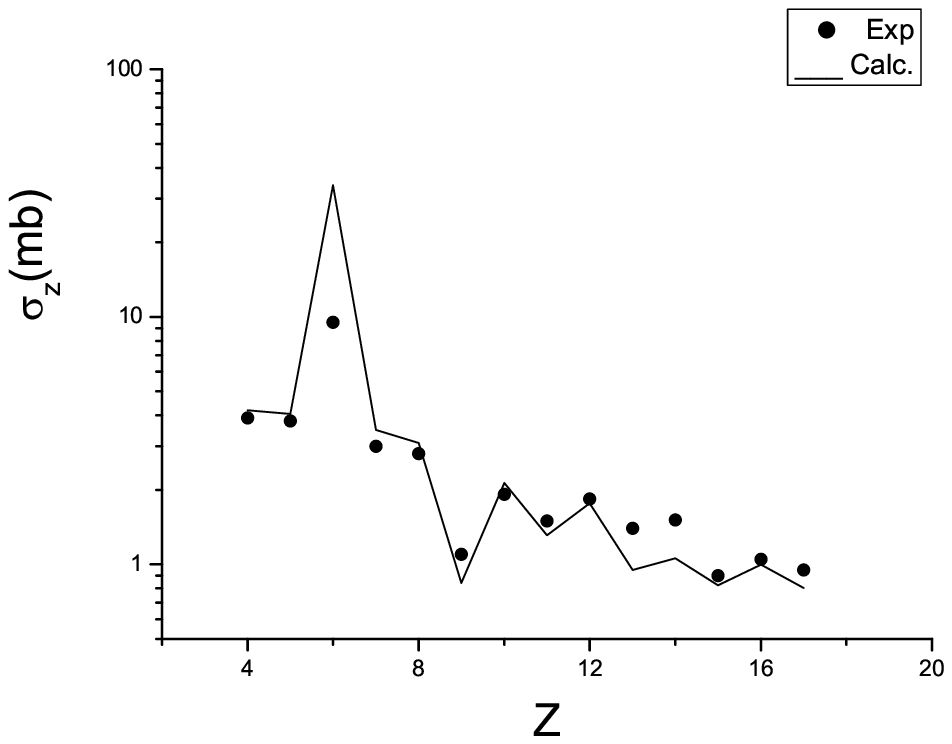}
\caption[Short caption for figure 2]{\label{63Cu12C} Comparison of calculated emission cross sections for the reaction
$^{63}$Cu+ $^{12}$C with initial bombarding energy $E=12.96$MeV/nucleon.}
\end{center}
\end{minipage}
\end{figure}

 The next reactions which we considered here are $^{78,}$  $^{86}$Kr $^{63}$Cu+
 $^{12}$C, which lead to CN $^{90,}$$^{98}$Mo\cite{Jing} and $^{75}$Br\cite{Han}. For the reaction $^{78}$ Kr + $^{12}$C, at initial bombarding energy  $8.52$ MeV/nucleon,$J^{kin}_{max}=39 \hbar$  and $J_{cr}=42 \hbar$, so one should take $J_{max}=39 \hbar$ in  calculating the cross sections.  The authors of this experimental work used in the fitting  of data with standard statistical code  $J_{max}=43 \hbar$. The results of calculations shown in Fig. \ref{78Kr12C}.  For the  reaction $^{86}$ Kr + $^{12}$C(Fig. \ref{86Kr12C}) with initial bombarding energy
 $E=9.31$ MeV/nucleon, $J_{cr}=45 \hbar$ and $J^{kin}_{max}=42 \hbar$,  and for the reaction $^{63}$Cu+
 $^{12}$C(Fig. \ref{63Cu12C}),  $J_{cr}=40 \hbar$ and $J^{kin}_{max}=48 \hbar$.
 One can observe from Figs. \ref{78Kr12C} and \ref{86Kr12C} that the emission cross sections for
 the heavier fragments are larger in more neutron deficient reactions. This can be explained that in neutron deficient
 reactions the neutron evaporation is suppressed, and the emission  of the complex fragments become more probable.
 Since we use experimental ground state binding energies in calculations, the odd-even effects are larger than in experiment.
 \section{Conclusions}
 The emission of complex fragments from excited compound nucleus is investigated within the DNS model.
 Transition state formalism was used to calculate decay probabilities for all channels. Formation of complex fragments is modeled as transition from CN  to DNS. The model was tested for asymmetric reactions with $A\sim 75-120$, where the angular momentums up to $\sim 45\hbar$ are involved. The results of calculations show good agreement with experimental data. And we are looking for further development of the model for heavier systems and for higher angular momentums.

 \section{Acknowledgments}
 We would like to acknowledge fruitful discussions with Prof. Wieleczko.
 The authors thank the support of the DFG (Bonn), FBR(Tashkent), RFBR (Moscow).

\end{document}